\documentclass[journal]{IEEEtran}
\usepackage{cite}
\usepackage{amsmath,amssymb,amsfonts}
\usepackage{algorithmic}
\usepackage{graphicx}
\usepackage{textcomp}
\usepackage{color}
\usepackage{subfiles}
\usepackage{float}
\usepackage{blindtext}
\usepackage{gensymb}

\usepackage{acronym}
\acrodef{AF}{Atrial Fibrillation}
\acrodef{SR}{Sinus Rhythm}
\acrodef{PVI}{Pulmonary Vein Isolation}
\acrodef{FIRM}{Focal Impulse and Rotor Modulation}
\acrodef{ECGI}{Electrocardiographic Imaging}
\acrodef{LAT}{Local Activation Time}
\acrodef{LA}{Left Atrium}
\acrodef{MRI}{Magnetic Resonance Imaging}
\acrodef{DT-MRI}{Diffusion Tensor Magnetic Resonance Imaging}
\acrodef{DE-MRI}{Delayed-Enhanced Magnetic Resonance Imaging}
\acrodef{LGE-MRI}{Late Gadolinium Enhancement Magnetic Resonance Imaging}
\acrodef{VT}{Ventricular Tachycardia}
\acrodef{CT}{Computed Tomography}
\acrodef{CV}{Conduction Velocity}
\acrodef{ICP}{Iterative Closest Point}
\acrodef{LSPV}{Left Superior Pulmonary Vein}
\acrodef{RSPV}{Right Superior Pulmonary Vein}
\acrodef{LIPV}{Left Inferior Pulmonary Vein}
\acrodef{RIPV}{Right Inferior Pulmonary Vein}
\acrodef{MV}{Mitral Valve}
\acrodef{BB}{Bachmann's Bundle}

\graphicspath{{figure/}}
\begin{document}

\title{Fiber Organization has Little Effect on Electrical Activation Patterns during Focal Arrhythmias in the Left Atrium}

\author{Jiyue He, Arkady M. Pertsov, Elizabeth M. Cherry, Flavio H. Fenton, Caroline H. Roney, Steven A. Niederer, Zirui Zang, Rahul Mangharam
\thanks{This work was supported in part by the National Science Foundation under grants CNS-1446675 (FHF), CNS-2028677,  CNS-1446312 (EMC), and CPS-1446664 (RM) Frontiers on Medical Cyber-Physical Systems, and in part by the National Institutes of Health under grant 1R01HL143450-01 (EMC and FHF).}
\thanks{Jiyue He (corresponding author, e-mail: jiyuehe@seas.upenn.edu), Zirui Zang, Rahul Mangharam: Department of Electrical and Systems Engineering, University of Pennsylvania, USA. Arkady Pertsov: Department of Pharmacology, Upstate Medical University, USA. Elizabeth Cherry: School of Computational Science and Engineering, Georgia Institute of Technology, USA. Flavio Fenton: School of Physics, Georgia Institute of Technology, USA. Caroline Roney: School of Engineering and Materials Science, Queen Mary University of London, UK. Steven Niederer: School of Biomedical Engineering and Imaging Sciences, King’s College London, UK.}
\thanks{Copyright (c) 2022 IEEE. Personal use of this material is permitted. However, permission to use this material for any other purposes must be obtained from the IEEE.}
\thanks{https://doi.org/10.1109/TBME.2022.3223063}
}

\maketitle

\begin{abstract}
Over the past two decades there has been a steady trend towards the development of realistic models of cardiac conduction with increasing levels of detail. However, making models more realistic complicates their personalization and use in clinical practice due to limited availability of tissue and cellular scale data. One such limitation is obtaining information about myocardial fiber organization in the clinical setting. In this study, we investigated a chimeric model of the left atrium utilizing clinically derived patient-specific atrial geometry and a realistic, yet foreign for a given patient fiber organization. We discovered that even significant variability of fiber organization had a relatively small effect on the spatio-temporal activation pattern during regular pacing. For a given pacing site, the activation maps were very similar across all fiber organizations tested. 
\end{abstract}


\section{Introduction}
Over the past two decades, there has been steady progress towards the development of realistic electrophysiological models \cite{o2011simulation, greene2022voltage, fenton2008models} of cardiac propagation with detailed description of gross cardiac anatomy \cite{Ho1999, Ho2009, iaizzo2016visible}, myocardial fiber organization \cite{Ho2001, Fastl2018}, and ionic currents involved in the generation of cardiac action potentials \cite{Lopez-Perez2015, berman2021interactive, kaboudian2019real}. At the same time, the focus of modeling studies has been gradually shifting from the investigation of fundamental mechanisms of cardiac arrhythmias \cite{weiss2002electrical, watanabe2001mechanisms, fenton1998fiber, groenendaal2014voltage, uzelac2017simultaneous} towards clinical applications \cite{arevalo2016arrhythmia, kayvanpour2015towards, niederer2020creation}. The range of applications that are actively been explored includes screening anti-arrhythmic \cite{Sanguinetti2003, bai2020silico} and pro-arrhythmic \cite{uzelac2021quantifying, mamoshina2021toward} drugs, optimizing anti-tachycardia \cite{Duncker2017} and anti-fibrillation \cite{ji2017synchronization, detal2022terminating} pacing protocols, and aiding arrhythmia ablation procedures \cite{Lim2020, prakosa2018personalized}, among others.

One of the key obstacles towards the development of constructing patient-specific models is not only speed in simulations \cite{kaboudian2019real} but adequate quantification, verification and information for such models \cite{pathmanathan2019comprehensive, pathmanathan2020data}. Detailed models have a very high-dimensional parameter space; moreover, most of the parameters cannot be measured on an individual basis and need to be postulated based on some generic values or recomputed libraries \cite{doste2019rule, bayer2012novel}. In this study, we investigate the problem of incorporating realistic myocardial fiber organization into an atrial conduction model, and their effect in the electrical propagation. 

The velocity of action potential propagation along fibers is typically 2-3 times faster than across fibers \cite{Valderrabano2007}, which makes fiber organization one of the key factors determining activation patterns in cardiac tissue \cite{franzone1993spread, fenton1998vortex}. Unlike atrial geometry, which can be readily obtained in a clinical setting via electroanatomical mapping \cite{Bhakta2008} or from clinical imaging, fiber organization cannot be obtained with a sufficient level of detail in live tissue \cite{Zhao2012}. To date, the best available resource for real patient fiber data is an ex-vivo \ac{DT-MRI} fiber database \cite{Pashakhanloo2016, Roney2021}, which required about 50 hours to scan each atrium \cite{Pashakhanloo2016}. 

The focus of our study is on modeling the left atrium, which is known to harbor the most common sources of atrial fibrillation and thus is particularly important with regard to potential clinical applications. Here we explore 3D chimeric models of the left atrium informed by a patient-specific \ac{DT-MRI}-derived geometry and various fiber direction patterns adopted from the \ac{DT-MRI} fiber database. Specifically, we investigate models with identical atrial geometry but with different fiber organization. By assuming one of these models as the ground-truth, we assess the capability of the others for predicting activation patterns produced by focal excitation sources located in different areas of the left atrium. 

\begin{figure*}[!ht]
\centering
\includegraphics[width = 1\textwidth]{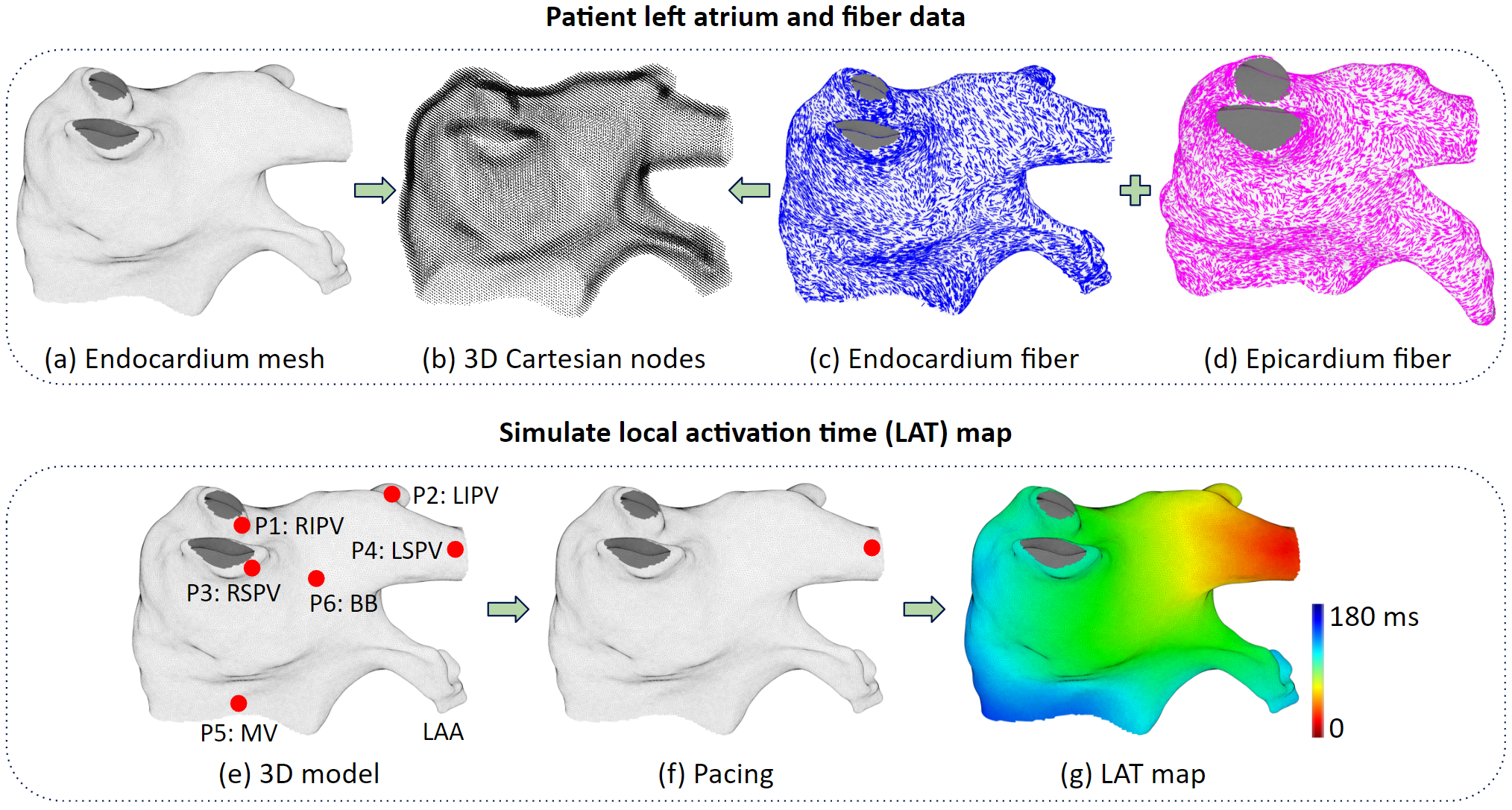}
\caption{Generation of surrogate activation maps in the ground-truth and chimeric models. (a) The endocardial mesh of left atrium 1 ($M_{endo1}$). (b) $M_{endo1}$ transformed into 3D Cartesian nodes to be used for modeling action potential propagation. (c, d) Endocardial and epicardial fiber organizations ($F_{endo1}$, $F_{endo2}$, …, $F_{endo7}$ and $F_{epi1}$, $F_{epi2}$, …, $F_{epi7}$) registered onto the nodes. (e) Locations of the pacing sites. (f, g) Example of a local activation time map produced from a pacing site. (RIPV: Right Inferior Pulmonary Vein, LIPV: Left Inferior Pulmonary Vein, RSPV: Right Superior Pulmonary Vein, LSPV: Left Superior Pulmonary Vein, MV: Mitral Valve, BB: Bachmann’s Bundle, LAA: Left Atrial Appendage.)}
\label{fig:flow chart}
\end{figure*}

The rest of this paper is organized as follows: We first describe the procedure for constructing in-silico heart models utilizing clinically derived patient-specific left atrial geometry and realistic fiber organization. We then show our results of heart models with different fiber organization in terms of \ac{LAT} and spatial activation patterns. In the discussion, we validate our models and explain the results in terms of a cancellation effect across the atrium. Finally, we describe the potential application to fiber-independent patient-specific heart models, list the limitations of our work, and summarize the main contributions.

\section{Method}
\subsection{Ground-Truth and Chimeric Models}
\label{sec:data source}

To generate ground-truth and chimeric models of action potential propagation in the left atrium, we used a \ac{DT-MRI} fiber database \cite{Roney2021}. It contains data sets from seven left atria, which are denoted as $LA_1$, $LA_2$, …, $LA_7$. The different fiber organizations are denoted as $F_1$, $F_2$, …, $F_7$, each can be subdivided into endocardium and epicardium fiber organizations, and are denoted as $F_{endo1}$, $F_{endo2}$…, $F_{endo7}$ and $F_{epi1}$, $F_{epi2}$, …, $F_{epi7}$, respectively. Endocardial and epicardial triangular meshes are denoted as $M_{endo1}$, $M_{endo2}$, …, $M_{endo7}$ and $M_{epi1}$, $M_{epi2}$, …,  $M_{epi7}$, respectively.

We used $M_{endo1}$ and $F_1$ to generate ground-truth models of action potential propagation and surrogate clinical activation maps. $M_{endo1}$ was chosen because $LA_1$'s fiber organization was found to be the optimal one, giving the least \ac{LAT} errors among the seven left atria for bi-layer model atrial fibrillation simulations \cite{Roney2021}. We used $M_{endo1}$ and $F_2$, $F_3$, …, $F_7$ to generate six chimeric models. We called them chimeric models because foreign fibers were registered onto the left atrium.

Fig. \ref{fig:flow chart} illustrates the generation of all models. The process involves generation of 3D Cartesian nodes around $M_{endo1}$, mapping of the respective fiber organizations onto the nodes, simulation of action potential propagation from stimuli in different locations, and generation of surrogate clinical activation maps. The latter were then used for evaluation of chimeric model performance in predicting activation patterns across different fiber patterns. 

\subsection{Atrium mesh and fiber processing}
\label{sec:mesh processing}
Triangular mesh $M_{endo1}$ has 63,112 vertices, 125,501 faces, and its average triangle edge length is 0.67 mm. The seven fiber organizations are registered onto $M_{endo1}$, resulting in seven different models (one ground-truth model and six chimeric models). This procedure requires morphing other meshes and fibers onto $M_{endo1}$. Details in Appendix \ref{app:fiber registration}. 

\begin{figure}[!ht]
\centering
\includegraphics[width = 0.5\textwidth]{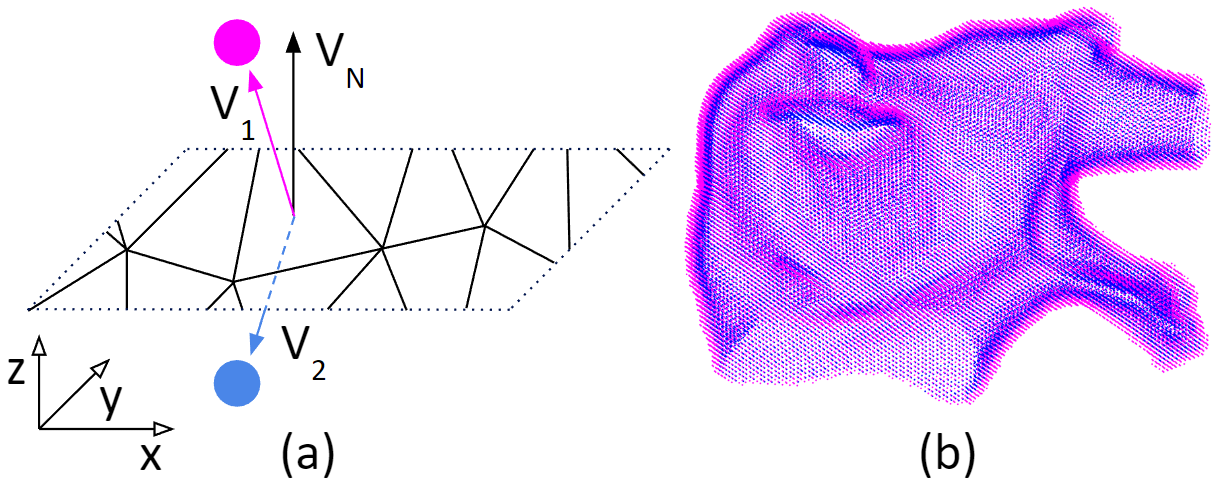}
\caption{Labeling of endocardial and epicardial nodes. (a) $V_1$ and $V_2$ are vectors pointing from the nearest triangle center to the respective node. $V_N$ is the surface normal vector. Because the angle between $V_1$ and $V_N$ is less than 90\degree, the magenta node was labeled as an epicardial node; If the angle is larger than 90\degree, the blue node was labeled as an endocardial node. If the angle is equal to 90\degree, then randomly assign it to endocardial or epicardial node. (b) All left atrial nodes labeled as epicardial (magenta) or endocardial (blue) nodes.}
\label{fig:endo epi node}
\end{figure}

The next step is to create 3D Cartesian nodes wrapped around $M_{endo1}$, as shown in Fig. \ref{fig:endo epi node}(b). First, a cubic space that contains $M_{endo1}$ is filled with equally spaced nodes (1 mm spacing), then nodes that are further than a distance threshold from the $M_{endo1}$ surface are removed. Here, the left atrium tissue thickness is set at 2 mm, which is in the range of clinically observed values \cite{Whitaker2016, Sun2018}, and the total number of Cartesian nodes is 60,691. The endocardium and epicardium layers split the tissue thickness in half, each having a thickness of 1 mm. The spatial resolution is chosen to be adequate (refer to Appendix \ref{app:resolution}) for accurate simulation without being computationally demanding. 

Each node of the grid was attributed to either the epicardial or endocardial layer, based on its location with respect to $M_{endo1}$. The appropriate layer was determined by the angle between the vector pointing to the respective node from the center of the closest triangle and the normal vector of this triangle. Note that although there are two normal directions (opposite to each other) for a triangle, we can easily determine which points out from the mesh: it is a property of a triangular mesh that the three vertices of a triangle are in a particular sequence, and the outward normal vector thus can be found according to the sequence and the right-hand-rule. Fig. \ref{fig:endo epi node} shows the labeled epicardial and endocardial nodes. Then the endocardial and epicardial fibers were assigned to these nodes via minimum distance mapping. 

\subsection{Heart model equations}
\label{sec:heart model equation}

To simulate action potentials, we used the Mitchell-Schaeffer model \cite{Mitchell2003}, as shown in \eqref{eq:mitchell schaeffer}. This model was chosen because its simplicity makes it efficient in 3D numerical simulations, and the parameters provide direct insight into changes in electrophysiological behavior. It models the inward current caused by sodium and calcium ion channels, outward current caused by potassium channels, and external stimulus current. 

\begin{align}
\label{eq:mitchell schaeffer}
\begin{split}
\frac{du}{dt}&=\frac{hu^2(1-u)}{\tau _{in}}-\frac{u}{\tau _{out}}+J_{stimulus}+\bigtriangledown \cdot (D\bigtriangledown u)
\\
\frac{dh}{dt}&=\left\{\begin{matrix}\ \frac{1-h}{\tau_{open}} \ \text{if}\ u<u_{gate} \\ \ \frac{-h}{\tau_{close}} \ \text{if}\ u \geq u_{gate}\end{matrix}\right.
\end{split}
\end{align}

The variables are as follows:
\begin{itemize}
\item $u$ is the transmembrane voltage and $h$ is an inactivation gating variable for the inward current.
\item $\tau_{in}, \tau_{close}, \tau_{out}, \tau_{open}$ and $u_{gate}$ are parameters that control the action potential shape.
\item $J_{stimulus}$ is an external current applied locally as impulses to initiate action potential. We specified this impulse to have 1 ms duration and a magnitude of 20.
\item $\bigtriangledown \cdot (D\bigtriangledown u)$ is the diffusion term, responsible for action potential propagation.
\end{itemize}

For each node, fiber anisotropy is introduced via a $3 \times 3$ diffusion tensor $D$ according to \eqref{eq:D}, 
\begin{align}
\label{eq:D}
D &= d \left ( rI+\left ( 1-r \right )ff^\top \right ) \\
r &= \frac{d_{T}}{d_{L}}= \left (\frac{CV_{T}}{CV_{L}}\right )^2
\end{align}
where $d$ is the diffusion coefficient. $r$ is the anisotropy ratio, a ratio of fiber's transverse to longitudinal diffusion coefficients, or the ratio of transverse to longitudinal conduction velocities squared. $I$ is the identity matrix, and $f$ is a $3 \times 1$ unit vector pointing along the fiber direction \cite{Elaff2018}.

The differential equations \eqref{eq:mitchell schaeffer} were solved using the explicit Euler method on the Cartesian nodes. We followed \cite{McFarlane2010} that assumed no-flux boundary conditions and used a 19-node stencil. The time step is 0.01 ms and the spatial resolution is 1 mm. Such resolution is adequate, refer to Appendix \ref{app:resolution} for details. The parameters are given nominal values as shown in Table \ref{tb:parameter} \cite{Cabrera2017, Roney2019}. With this setting, the \ac{CV} will be around 0.69 m/s, which is a typical value for the atrium. The model is implemented in MATLAB (MathWorks, Natick, Massachusetts, United States) and accelerated with GPU computing using CUDA kernels (Nvidia, Santa Clara, California, United States). 

\begin{table}[!ht]
    \centering
    \caption{Nominal parameter values}
    \begin{tabular}{ ccccccc }
    \hline
        $\tau_{in}$ & $\tau_{out}$ & $\tau_{open}$ & $\tau_{close}$ & $u_{gate}$ & $r$ & $d$ \\\hline
        0.3 ms & 6 ms & 120 ms & 150 ms & 0.13 mV & 0.2 & 1 \\ 
    \hline
    \end{tabular}
    \label{tb:parameter}
    \begin{flushleft}
    \end{flushleft}
\end{table}

\section{Results}
\subsection{Fibers vary significantly across different atria}
To compare fiber organization in different models of the left atrium, all fiber patterns were registered onto the same mesh ($M_{endo1}$). To compare the different fibers, we made a reference frame transformation, details refer to Appendix \ref{app:compare fiber}. The comparison was done separately for the endocardial and epicardial fibers. The correlations of fiber orientations are in the range between -0.12 and 0.18 (with the exception of auto-correlation). Low correlation indicates significant variation in fiber organization across different  atria, which is consistent with earlier observations \cite{Ho2001, Pashakhanloo2016}.

\subsection{Fibers vary within the left atrium}
\begin{figure}[!ht]
\centering
\includegraphics[width = 0.48\textwidth]{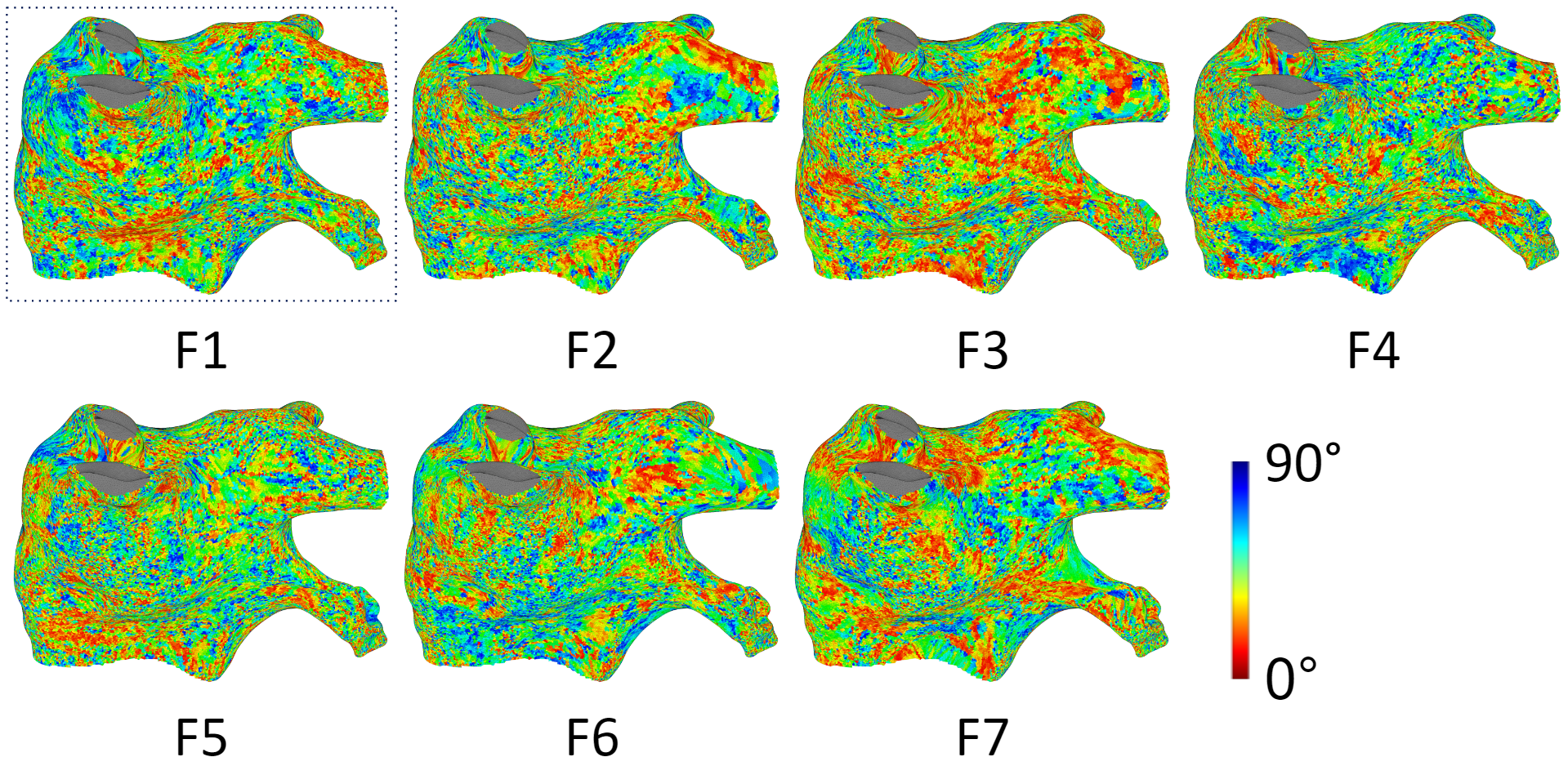}
\caption{Local differences between endocardial and epicardial fiber orientations ($\Delta\theta$ maps) in the ground-truth F1 (box on the top left panel) and chimeric models of the left atrium (F2 to F7). Red and blue colors represent small and large $\Delta\theta$, respectively. Random local color variations indicate lack of large regions with the same $\Delta\theta$ value.}
\label{fig:delta theta map}
\end{figure}

Importantly, we found that in all atria the fiber pattern of the endocardium is very different from that of the epicardium. We define $\Delta\theta$ as the fiber angle difference between endocardial and epicardial fibers in a given location. Fig. \ref{fig:delta theta map} shows the $\Delta\theta$ maps in all seven left atria analyzed in this study; red represents small and blue represents large $\Delta\theta$. No large regions have the same $\Delta\theta$ value, and there is no regularity in the spatial distribution of the small and large $\Delta\theta$ regions in all atria tested.

Table \ref{tb:delta theta percentage} shows the relative area occupied by smaller $\Delta\theta$ regions ($\Delta\theta \leq 45^{\circ}$) in different atria. Areas with smaller $\Delta\theta$ manifest higher effective anisotropy, which contributes to increased sensitivity of activation patterns to fiber organisation.  

\begin{table}[!ht]
    \centering
    \caption{$\Delta\theta$ area analysis}
    \begin{tabular}{ c c c c c c c c}
    \hline
F1 & F2 & F3 & F4 & F5 & F6 & F7 \\ \hline
57\% & 56\% & 65\% & 55\% & 57\% & 53\% & 64\% \\ \hline 
    \end{tabular}
    \label{tb:delta theta percentage}
    \begin{flushleft}
    Percentage of the atrium area occupied by regions with $\Delta\theta \leq 45^{\circ}$ in different models. 
    \end{flushleft}
\end{table}
\subsection{Fibers do not significantly affect activation pattern}
\label{sec:42 simulations}
\subsubsection{Local activation time}
To test the effects of fiber organization on activation patterns, we generated 42 simulations: six pacing sites each with seven different fiber patterns. The six pacing sites, shown in Fig. \ref{fig:flow chart}(e), are P1: \ac{RIPV}, P2: \ac{LIPV}, P3: \ac{RSPV}, P4: \ac{LSPV}, P5: \ac{MV}, and P6: \ac{BB}.

\begin{figure}[!ht]
\centering
\includegraphics[width = 0.48\textwidth]{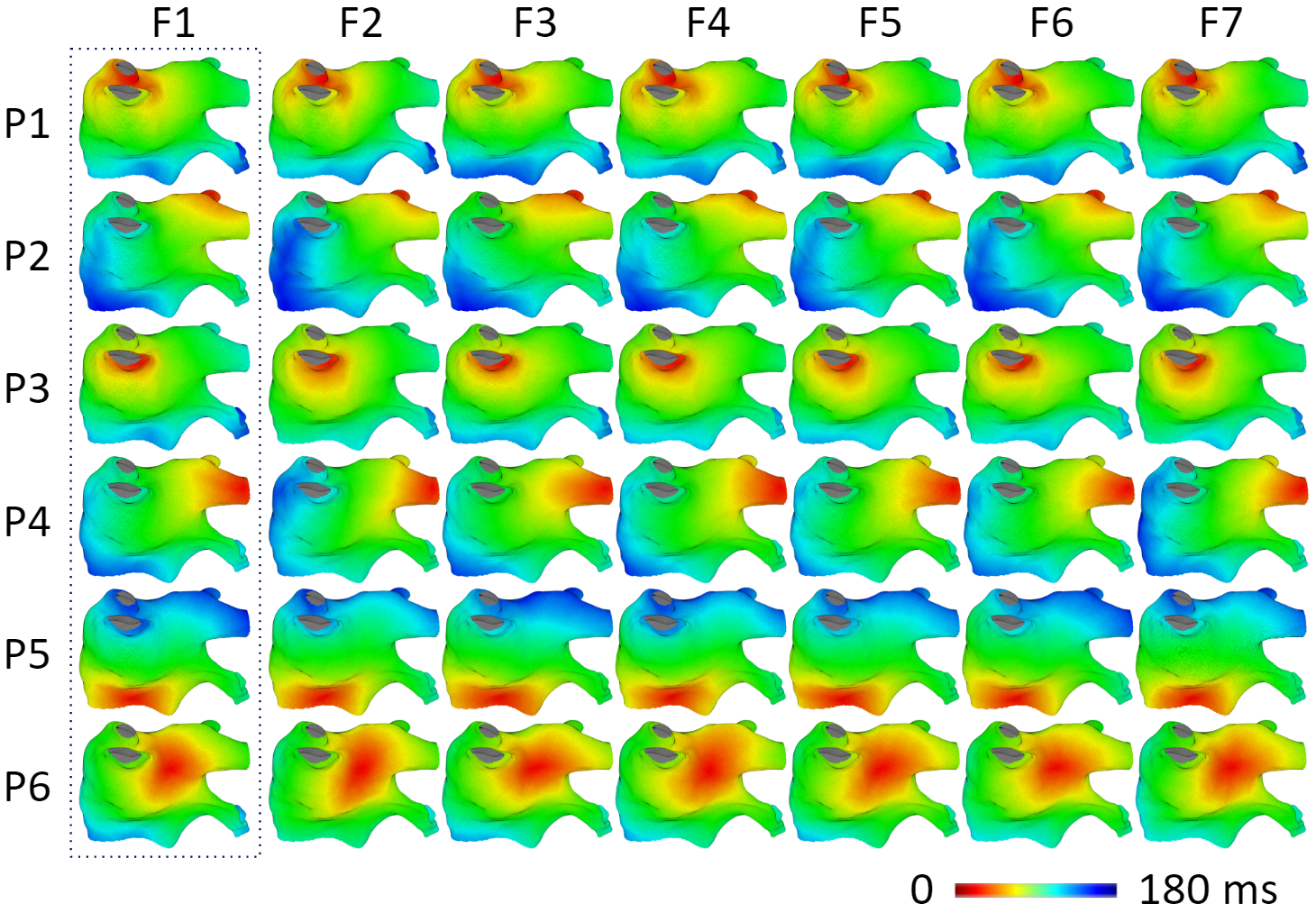}
\caption{\ac{LAT} maps produced by pacing from six different locations in the ground-truth (column F1) and chimeric models (columns F2 to F7). Color scales are normalized to each individual map. The maps in each row are similar to each other, demonstrating that despite significant differences in fiber patterns the activation pattern is not significantly affected.}
\label{fig:lat maps}
\end{figure}

The resulting \ac{LAT} maps are shown in Fig. \ref{fig:lat maps}. All maps are on $M_{endo1}$. The first column shows the simulation results in the ground-truth model built upon the intrinsic left atrium geometry and the fiber $F_{endo1}$ and $F_{epi1}$. Columns 2-7 are \ac{LAT} maps obtained in models in which the intrinsic fiber was substituted with other fibers. The \ac{LAT} maps of each row are similar, indicating that differences in fiber organization do not significantly affect the activation pattern.

To compare quantitatively the activation maps generated using the chimeric models with the ground-truth we calculated \ac{LAT} errors (see Table \ref{tb:lat err abs}). The average \ac{LAT} error is 7.8 ms, which is relatively small compared to the time it takes for the activation to travel through the entire left atrium (4.3\% of 180 ms). The LAT correlation is also quite high ranging between 0.94 and 0.99 for different pacing sites and across all models. Also, we noticed that average \ac{LAT} error (Table \ref{tb:lat err abs} bottom row) correlates with the relative size of the region with small $\Delta\theta$ (Table \ref{tb:delta theta percentage}). 

\begin{table}[!ht]
    \centering
    \caption{absolute LAT error (ms)}
    \begin{tabular}{ c | c c c c c c | c}
    \hline
         & F2 & F3 & F4 & F5 & F6 & F7 & \textbf{Avg}\\ \hline
P1 & 4.6 & 6.1 & 5.5 & 4.2 & 5.7 & 5.0 & \textbf{5.2} \\ 
P2 & 6.0 & 9.7 & 4.0 & 4.9 & 5.7 & 6.3 & \textbf{6.1} \\ 
P3 & 9.0 & 11.5 & 7.1 & 8.8 & 10.4 & 12.5 & \textbf{9.9} \\ 
P4 & 12.1 & 9.3 & 7.0 & 9.7 & 6.9 & 9.5 & \textbf{9.1} \\ 
P5 & 13.8 & 7.5 & 13.6 & 3.8 & 6.7 & 7.8 & \textbf{8.9} \\ 
P6 & 7.2 & 8.1 & 7.5 & 6.5 & 6.6 & 8.4 & \textbf{7.4} \\ \hline 
\textbf{Avg} & \textbf{8.8} & \textbf{8.7} & \textbf{7.5} & \textbf{6.3} & \textbf{7.0} & \textbf{8.3} & \textbf{7.8} \\ \hline 
    \end{tabular}
    \label{tb:lat err abs}
    \begin{flushleft}
    The last row shows the average LAT errors across all pacing sites in different fiber models, the last column shows the average error for a given pacing site across different models. The overall average is 7.8 ms or 4.3\% of the LAT range (which is 180 ms).
    \end{flushleft}
\end{table}

\subsubsection{Latest activation location}
\begin{figure*}[!ht]
\centering
\includegraphics[width = 1\textwidth]{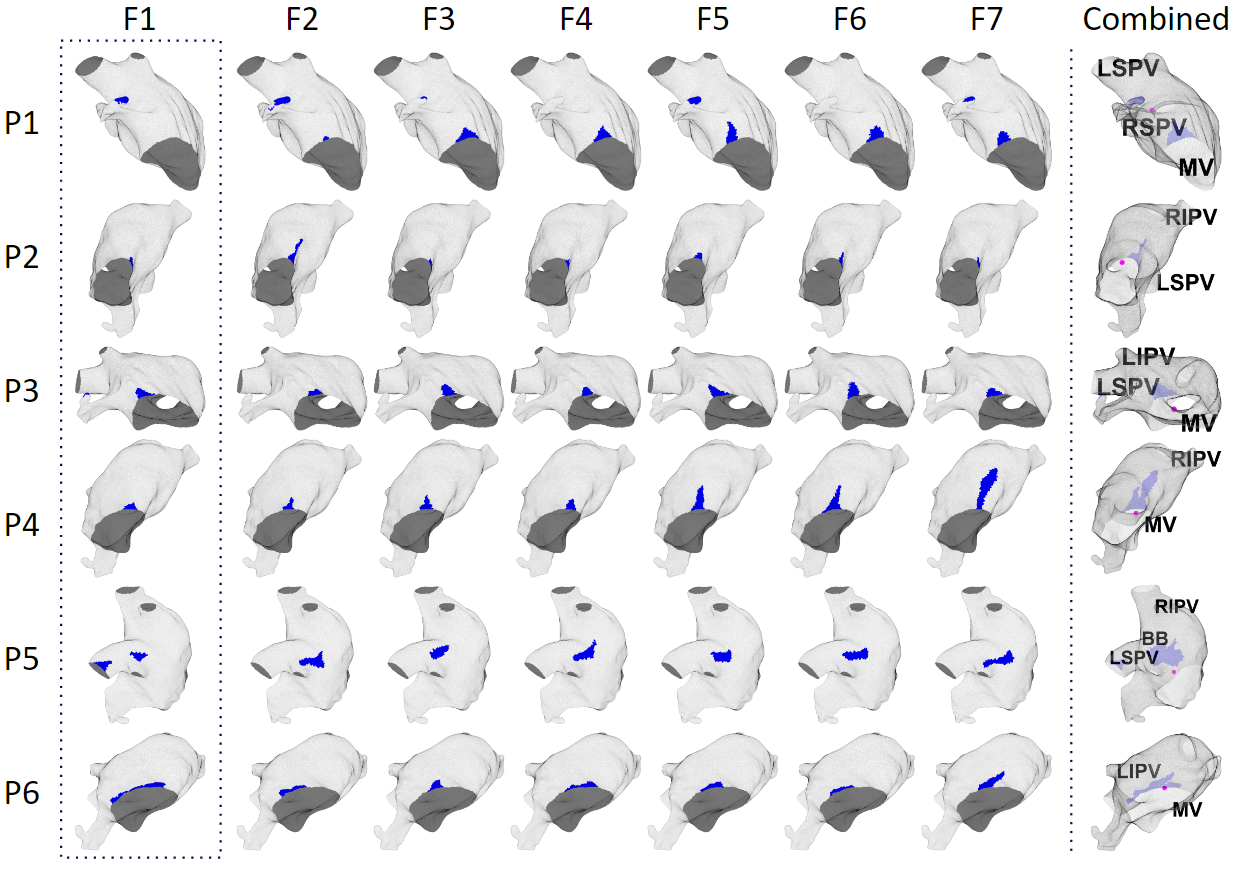}
\caption{The regions of the latest activation derived from the activation maps shown in Fig. \ref{fig:lat maps}. The view angles are adjusted for each of pacing sites for better visualization. The first column (boxed) shows the results in the ground-truth model. Blue regions show the sites with the \ac{LAT} values within 5 ms from the maximum. The average maximum \ac{LAT} of the 42 scenarios is 180 ms. The last column shows the superposition of the latest activation regions for all fiber organizations. The magenta dots show the pacing sites on the back side of the atrium. For a given pacing site, the position of the blue regions are reasonably well preserved across different fiber organizations.}
\label{fig:latest activation time locations}
\end{figure*}

Fig. \ref{fig:latest activation time locations} shows the latest activation locations in the left atrium. Blue regions indicate where \ac{LAT} values are within 5 ms of the maximum LAT; The first column shows the results obtained in the ground-truth model with the intrinsic fiber pattern. Other columns are results obtained with different fiber organizations. Notably, in most of cases, the regions of late activation have similar locations, indicating that different fiber organizations do not alter the start-to-end activation wave patterns significantly. For example, rows P3 and P6 have consistent latest activation locations regardless of fiber organization; rows P2 and P4 show some instances with dilated regions, and row P1 has some different locations, but the clusters are still on the opposite side of the pacing site, which means the general activation pattern remains the same. 

\begin{table}[!ht]
    \centering
    \caption{latest activation location difference (mm)}
    \begin{tabular}{ c | c c c c c c | c}
    \hline
         & F2 & F3 & F4 & F5 & F6 & F7 & \textbf{Avg} \\ \hline
P1 & 3.37 & 34.63 & 38.75 & 16.80 & 38.30 & 24.17 & \textbf{26.00} \\ 
P2 & 6.95 & 1.86 & 1.73 & 8.02 & 1.38 & 3.17 & \textbf{3.85} \\ 
P3 & 17.28 & 14.10 & 15.83 & 10.72 & 10.60 & 12.62 & \textbf{13.52} \\ 
P4 & 2.95 & 3.16 & 4.24 & 6.13 & 6.32 & 17.12 & \textbf{6.65} \\ 
P5 & 26.09 & 18.61 & 26.81 & 26.09 & 23.61 & 27.40 & \textbf{24.77} \\ 
P6 & 9.72 & 4.68 & 5.23 & 2.49 & 8.33 & 5.78 & \textbf{6.04} \\ \hline 
\textbf{Avg} & \textbf{11.06} & \textbf{12.84} & \textbf{15.43} & \textbf{11.71} & \textbf{14.76} & \textbf{15.04} & \textbf{13.47} \\ \hline 
    \end{tabular}
    \label{tb:latest activation location difference}
    \begin{flushleft}
    Latest activation location difference is the distance between the center of the blue area for F2-F7 and F1 (the truth). The mean is 13.47 mm and the median is 9.43 mm, which are relatively small compared to the left atrium size, indicating that the latest activation locations for different fiber organizations are similar. Note that the values in rows P5 and P6 are very different, which demonstrates that the latest activation location differences depend on pacing location.
    \end{flushleft}
\end{table}

The quantitative analysis is summarized in Table \ref{tb:latest activation location difference}. For each simulation illustrated in Fig. \ref{fig:latest activation time locations} columns F2-F7, we compute the distance of the center of the blue area to the truth. For most scenarios, the distances are small (median: 9.43 mm; average: 13.47 mm) compared to the size of the left atrium mesh $M_{endo1}$ (84 mm $\times$ 97 mm $\times$ 88 mm). Notably, the average across different pacing sites  does not change significantly from model to model (see the last row in Table \ref{tb:latest activation location difference}). Yet, for some pacing sites, they are consistently better than for other pacing sites, for example, compare rows P5 and P6.

\subsection{Fibers have local effects on activation propagation}
\begin{figure*}[!ht]
\centering
\includegraphics[width = 1\textwidth]{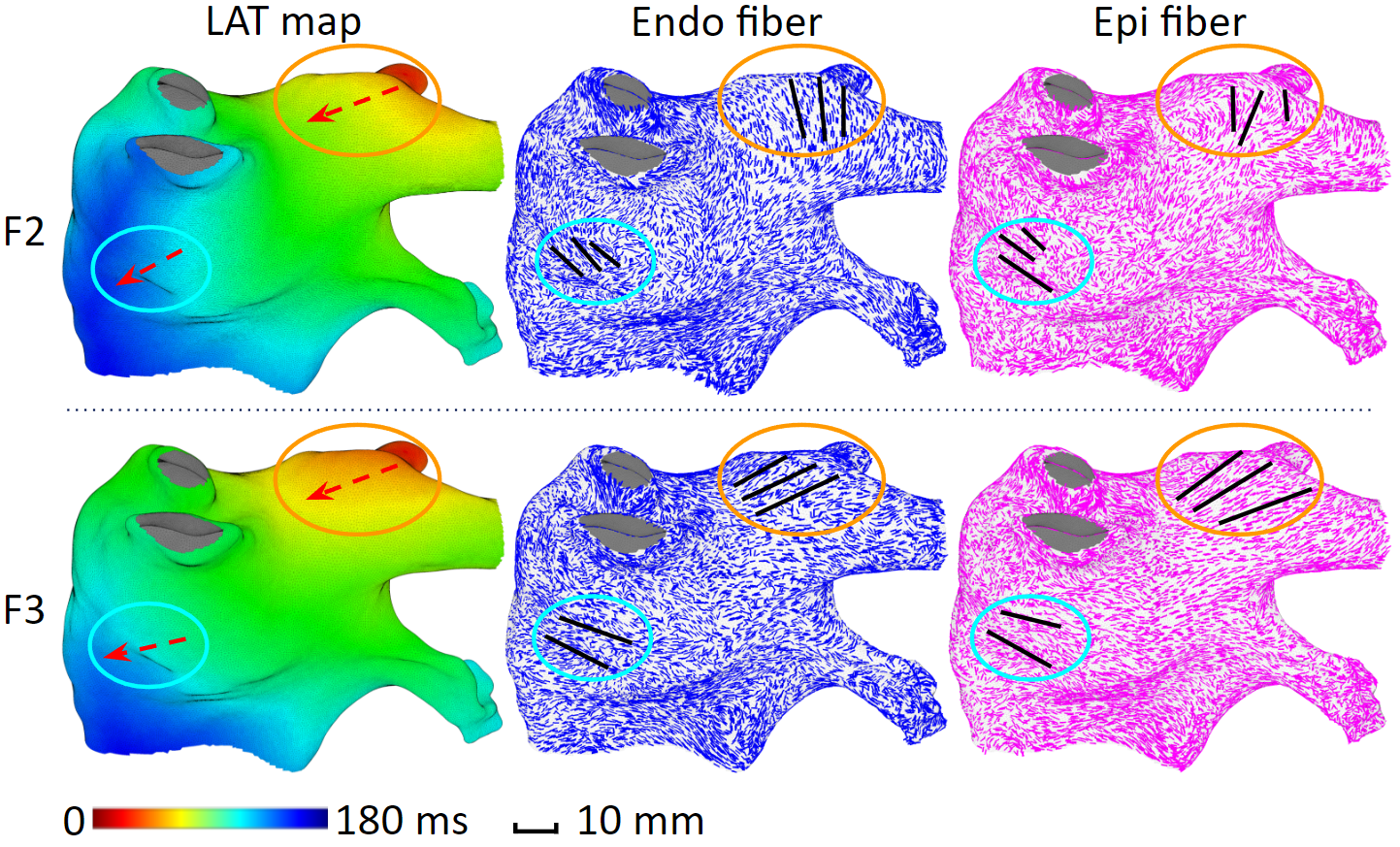}
\caption{Local effect of fiber organization. Upper and lower panels show LAT maps (left) and fiber organization (center and right) in models F2 and F3, respectively. Red arrows: propagation direction. Black lines: prevailing fiber direction in the areas of interest (orange and cyan ovals). The LAT gradient inside the orange oval is conspicuously greater in F2 than in F3 (red-green vs red-yellow). This is because the endocardial and epicardial fibers in F2 are perpendicular to the activation direction, resulting in slower propagation. This is not the case for F3 where propagation is parallel to fibers. Such local effects exist throughout the entire atrium, but they do not alter the overall activation pattern much.}
\label{fig:local effect}
\end{figure*}

Although differences in fiber organization have little effect on the large scale, they can be observed locally. Fig. \ref{fig:local effect} shows activation maps generated using models F2 and F3 paced at P2. The activation propagates in the direction of the red-dashed arrow. In the F2 row, the activation propagation slows down in the orange circled area (it has a smaller red region) because the fiber direction is perpendicular to the activation direction. In the F3 row, because fiber direction is along the activation direction, it accelerates propagation and results in a larger red area. Such acceleration and slow-down effects occur throughout the left atrium. The cyan-circled area is another example: propagation is slower in the F2 row than F3 row, resulted in more blue area.

\subsection{Fiber effects with different anisotropy ratios}
\begin{figure*}[!ht]
\centering
\includegraphics[width = 1\textwidth]{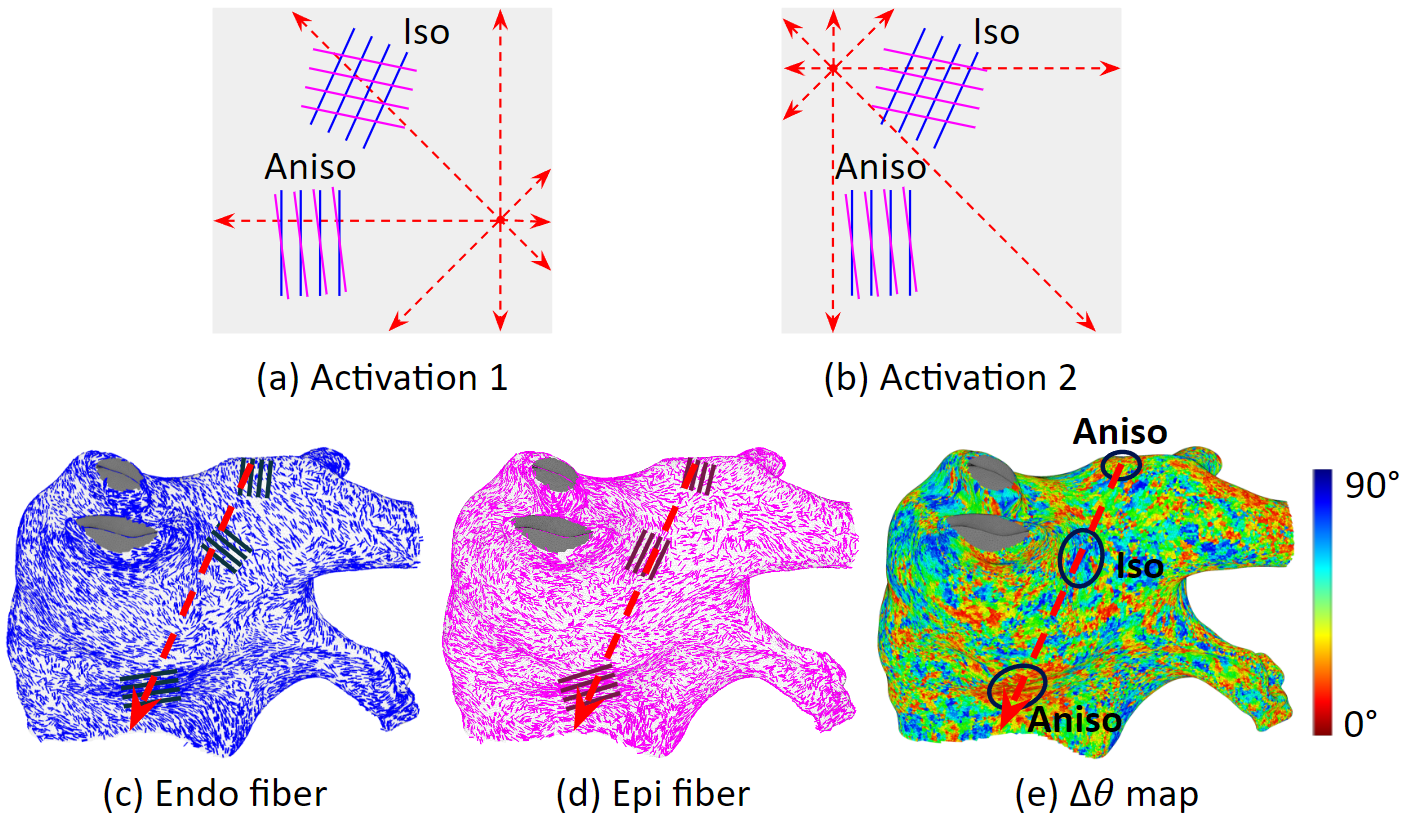}
\caption{The cancellation effect explained via the fibers: there is no macro level consistent fiber organization, therefore local fiber effects cancel each other. (a, b) show fiber’s local effect. Blue lines represent endocardium fiber, magenta lines represent epicardium fiber. Aniso region has a small $\Delta\theta$ (about 0\degree), and Iso region has a large $\Delta\theta$ (about 90\degree). Red arrows are activation wave traveling directions. For activation scenario (a), The Aniso region has a slow down effect on the activation; However, for activation scenario (b), The Aniso region has a speed up effect on the activation. And for both (a) and (b), the propagation speed in the Iso region are similar. (c, d, e) show that the local effects cancel each other at the macro level. (c) is endocardium fiber, (d) is epicardium fiber, and (e) is $\Delta\theta$ map. Along the activation path marked by the red dashed arrow, the propagation speed increases in the top Aniso region, then it goes trough an Iso region that does not change the speed much, finally it enters the bottom Aniso region and the speed decreases. In summary, activation speed first increases then decreases, therefore the effects cancel each other.}
\label{fig:cancellation effect}
\end{figure*}

\begin{table}[!ht]
    \centering
    \caption{Different anisotropy ratios}
    \begin{tabular}{ c | c c c }
    \hline
        $r$ & 0.1 & 0.2 & 0.5 \\\hline
        LAT Error (ms) & 10.6 or 5.5\% & 7.8 or 4.3\% & 3.15 or 2.0\% \\ 
    \hline
    \end{tabular}
    \label{tb:r}
    \begin{flushleft}
    The percentage is with respect to the time it takes the activation to travel through the left atrium. For r = 0.1, 0.2, 0.5, that time is 192, 180, 155 ms respectively. 
    \end{flushleft}
\end{table}

The majority of simulations of atrial propagation use values of $r$ between 0.1 and 0.2. Some publications use $r = 0.11$ \cite{Coster2018, Aslanidi2011}, while some others use $r = 0.2$ \cite{Boyle2019}. To investigate the robustness of our conclusions, we performed the same experiment as described in Section \ref{sec:42 simulations} with $r = 0.1$ and $r = 0.5$, and the results are summarized in Table \ref{tb:r}. The small errors suggest that the anisotropy ratio is not the main factor responsible for the low sensitivity of the large scale activation patterns to specific fiber organization.

\subsection{Uncouple the two fiber layers}
The reason for fiber organization having little effect on the activation patterns at the large scale could be partly due to significant differences in fiber orientation on the epicardial and endocardial layers (see Fig. \ref{fig:delta theta map}) which cause reduction in apparent anisotropy. But the left atrial wall was not always two layers in all parts \cite{Ho2009, Ho2012}. To evaluate the contribution of this effect on the large scale activation we generated seven models in which epicardial layers were assigned the endocardial fiber orientations, thus producing identical fiber orientation in both layers, which is equivalent to completely uncouple the endocardial layer from the epicardium. 

We hypothesized that this modification should amplify the apparent anisotropy, and consequently the differences between activation patterns in different models. Using these modified models, we performed the same experiment as described in Section \ref{sec:42 simulations} and compared the results. It might be expected, the errors in  \ac{LAT} in the models with identical epicardial and endocardial fiber orientations was greater than in the original models, however, the difference was not very large (9.1 ms vs 7.8 ms). 

This suggests that the difference in the endocardium and epicardium fiber orientations is not the main factor responsible for low sensitivity of the large scale activation patterns to specific fiber organization.

\section{Discussion}
\subsection{Contemporary heart modeling with fibers}
Electrophysiological heart modeling is becoming increasingly realistic with an aim to guide atrial fibrillation ablation procedures. One important element of a heart model is the fiber organization. However, accurate and high-resolution fiber data is not available for clinical use. The major difficulty of obtaining such in-vivo fiber organization is the time it takes to scan the fibers. The current \ac{DT-MRI} technology can scan high-resolution fibers in about 50 hours \cite{Pashakhanloo2016}, which is clinically impractical. 

One way to quickly obtain fiber organization is to mathematically calculate synthetic fibers based on atrial geometry \cite{Fastl2018, Krueger2011, Labarthe2021, Wachter2015, Saliani2021}. However, these fibers are artificial and do not represent the truth. Another option is to register foreign fibers. It has been found that one certain patient's atrial fiber organization can be generalized to many different patients \cite{Roney2021}. 

In this paper, we tried an alternative way, we ask the question: how big an effect does fiber organization have on activation patterns? If the effects are small, then we could create a clinically practical heart model that does not involve fiber data.

\subsection{Cancellation effect}
Our data suggest that the effects of fiber organizations are cancelled because fiber organizations vary across the left atrium. This cancellation effect can be explained at two levels. 

At the micro level, depending on the location of the activation origin, wave propagation can be shaped by the local fiber organization. As shown in Fig. \ref{fig:cancellation effect}(a, b), they represent the same tissue region but has different activations. Aniso region has small $\Delta\theta$ (about 0\degree), it is highly anisotropic, and have the strongest local effect on activation propagation; Iso region has large $\Delta\theta$ (about 90\degree), it is highly isotropic, and have the least effect on activation propagation. For Iso region, activation waves will travel through them in a similar manner regardless of the direction of travel. For Aniso region, depends on the direction of the activation wave, its travelling speed can be decreased (as shown in (a)) or increased (as shown in (b)). 

At the macro level, the local effects from the micro level will be cancelled by each other. As shown in Fig. \ref{fig:cancellation effect}(c, d, e), along a global path, the activation wave can travel through areas that increase and areas that decrease propagation speed, resulting in an overall zero effect. 

\begin{figure}[!ht]
\centering
\includegraphics[width = 0.5\textwidth]{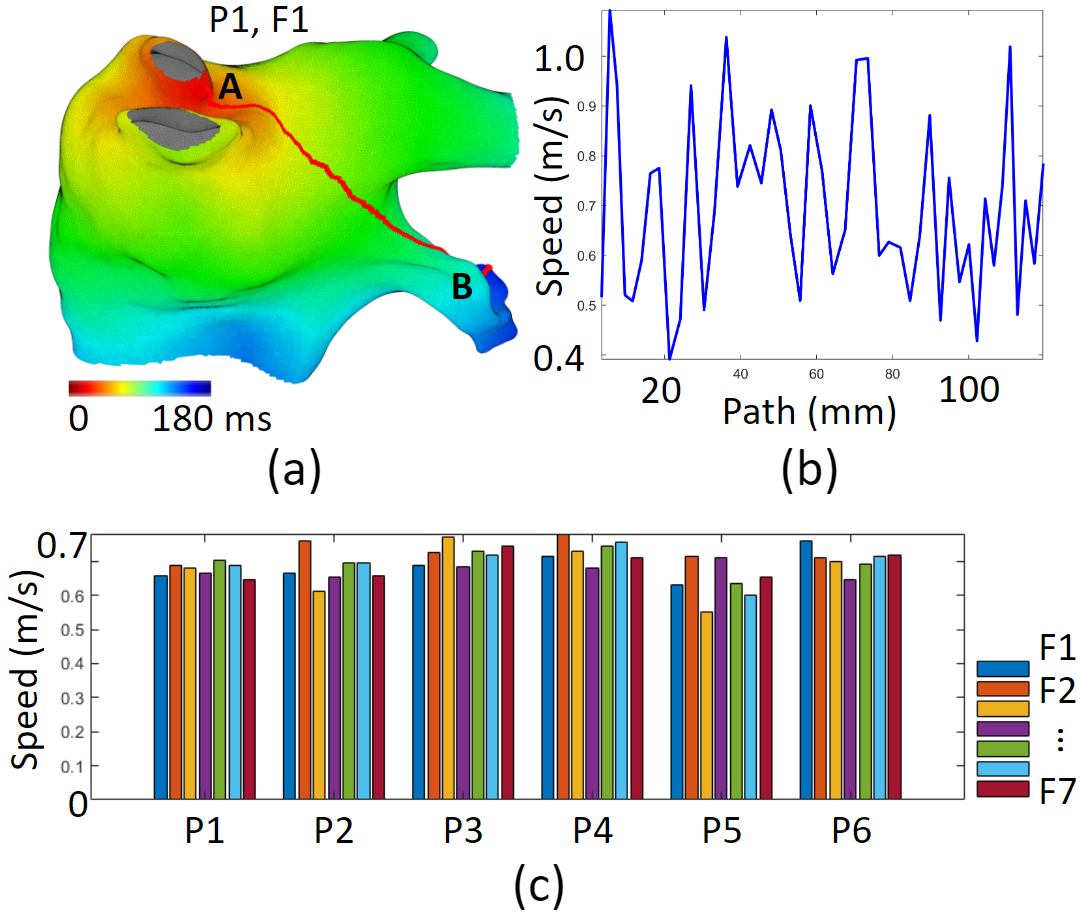}
\caption{An example of propagation speed along a path. (a) The path (red line) is the geodesic line from the earliest (point A) to the latest activation location (point B). (b) The path’s value is the geodesic length from the point along the path to point A. We can see that the propagation speed increases and decreases along the path. (c) The bar plots are the average propagation speed. We can see that for all the six different pacing scenarios, the average propagation speed values are similar regardless of which fiber organization was implemented. (Note that propagation speed is not necessarily equal to \ac{CV}, because \ac{CV} is in the \ac{LAT} gradient direction while the path is not necessarily aligned with that. We did not compute \ac{CV} along the path because it is not easy to get accurate results \cite{Cantwell2015TechniquesFA}.)}
\label{fig:cv}
\end{figure}

Quantitatively, the propagation speed along a path is shown in Fig. \ref{fig:cv}. The path is the geodesic line (minimum distance path) from the activation origin (point A) to the latest activation location (point B) as shown in (a), and it can be found via Dijkstra's algorithm. The path has a length of about 120 mm. The propagation speed of a point along the path is calculated as distance divided by time, where distance is the geodesic length from that point to point A, and the time is the \ac{LAT} values difference between that point and point A. We can see that the propagation speeds vary around its average value along the path as shown in (b), and the average propagation speeds are similar regardless of fiber organizations as shown in (c). Such phenomenon happened for all 42 scenarios. Because of this phenomenon, the fiber organization’s local effect does not accumulate along the path, resulting in a small macro effect.

\subsection{Potential applications: fiber-independent model}
In our previous research \cite{He2021}, we showed that we can construct a model of the left atrium to reproduce clinical electroanatomical mapping data \cite{He2019} without incorporating fiber organization. We validated our fiber-independent model with data from 15 patients and the performance was good: the average absolute \ac{LAT} error was 5.47 ms for sinus rhythm and 10.97 ms for flutter and tachycardia, and the average correlation was 0.95 for sinus rhythm and 0.81 for flutter and tachycardia. There are also other studies support the use of heart model without fibers \cite{Virag2002, Ruchat2007, Gray1996}.

In this work, because we found that fibers did not significantly affect activation patterns, we expect that the fiber-independent model can be useful for predicting fibrillation source locations. Because the fiber-independent model only requires heart geometry and electrograms, it could be integrated into contemporary electroanatomical mapping systems to provide real-time atrial fibrillation ablation guidance.

\section{Limitations}
We found that fiber organization does not significantly affect activation patterns. However, there are important limitations to the scenarios we considered that informed this finding. 

1) We focused only on sustained sources, such as focal atrial tachycardia, which may occur following atrial fibrillation ablation. It is possible that non-stationary sources such as drifting rotors could be affected by fibers. 

2) We did not consider scars, which could play an important role in atrial fibrillation dynamics \cite{Gonzales2014}. 

3) The Mitchell-Schaeffer model we used is not a detailed ionic model, and it may not be a good choice to model complex rhythms such as atrial fibrillation. Still such a two-component model is good for modeling periodic propagation. We utilized the computationally more efficient mono-domain model. It is well established that more detailed bi-domain models are required for accurate simulation of electrical activity in the immediate vicinity of the stimulating electrodes and for modelling electrical defibrillation \cite{Roth2021}. With regard to accuracy, the bi-domain models have practically no advantages over mono-domain models for simulations of propagation patterns of external stimulus \cite{Potse2006}.

4) We simplify fiber organization into only two layers. The real left atrium has many more layers, and the number of layers also varies in different regions, as does the atrial thickness. If more layers were incorporated, we would need to study if the effects of fibers would become stronger as well as whether the fiber direction changes abruptly or gradually through the thickness. 

5) We examined fiber organizations of the left atrium, because the most common atrial fibrillation sources are in the left atrium; therefore, it has more available clinical electroanatomical mapping data than the right atrium. We have not examined whether our findings would hold in the right atrium.  

\section{Conclusion}
In this paper, we found that 1) Fiber organization varies significantly across different left atria and within a left atrium. 2) Fibers have local effects on activation propagation but such local effects cancel each other at the macro level. And 3) fibers do not significantly affect activation pattern.

In summary, for left atrial focal arrhythmia, we found that the global activation patterns do not seem to be significantly affected by fiber organization. Therefore, fiber organization may not be essential for accurate heart modeling of arrhythmias in  the left atrium, and thus  more practical heart models for some clinical applications could be fiber-independent models. It has been shown that there is an asynchrony and dissociation between activation on the epicardium and endocardium during atrial  arrhythmias \cite{de2016direct, eckstein2013transmural, verheule2014role}; these and our results further suggest that it may be more important to consider the role of the complex trabecular network \cite{verheule2014role, wu1998role, berman2021interactive} during arrhythmias.

\appendices
\begin{appendices}

\section{Fiber registration}
\label{app:fiber registration}

\begin{figure}[!ht]
\centering
\includegraphics[width = 0.5\textwidth]{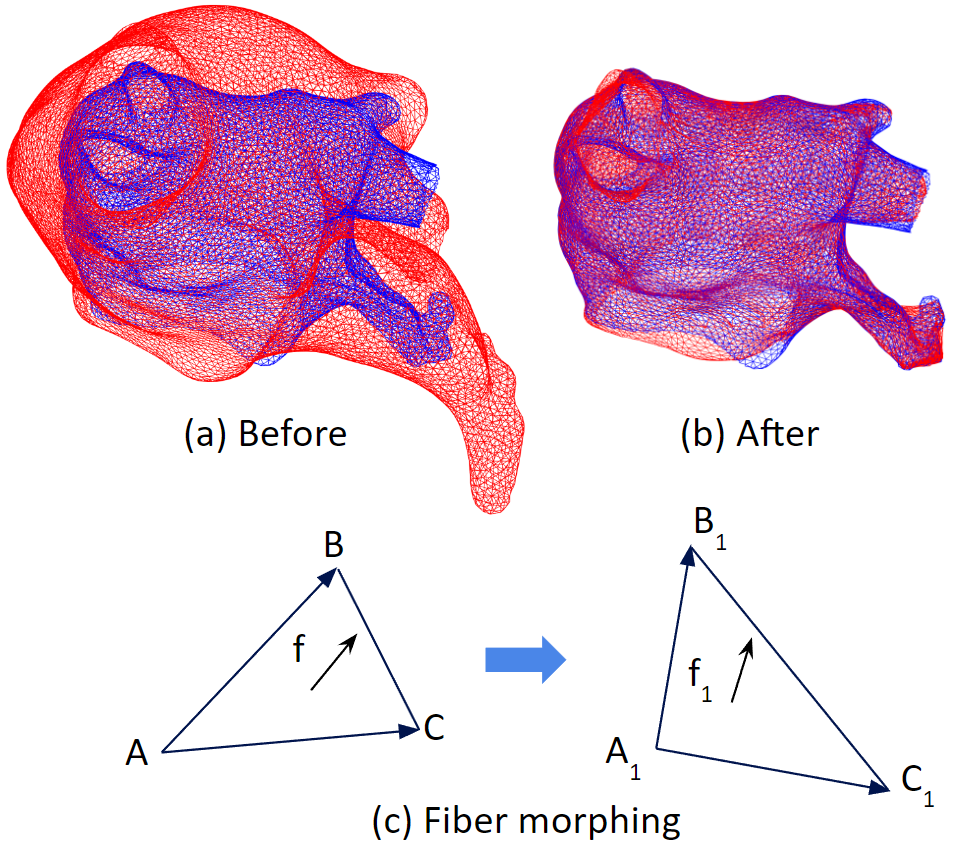}
\caption{Fiber registration. (a) and (b) are the before and after mesh morphing. (c) shows how to transform fiber from the original mesh (red mesh in (a)) to the morphed mesh (red mesh in (b)): Given xyz coordinates of $A$, $B$, $C$, $A_1$, $B_1$, $C_1$, and the fiber (3D vector $f$), calculate $f_1$. We have $f = a \textbf{AB} + b \textbf{AC}$, by solving for $a$ and $b$, we can obtain $f_1 = a \textbf{A}_1\textbf{B}_1 + b \textbf{A}_1\textbf{C}_1$, then make $f_1$ unit length, it will be the morphed fiber. (Note that $f = a \textbf{AB} + b \textbf{AC}$ will give 3 equations, since $\textbf{AB}$, $\textbf{AC}$ and $f$ are co-plane, any 2 of the 3 equations will give the same $a$ and $b$.)}
\label{fig:fiber registration}
\end{figure}

The seven endocardium and seven epicardium fibers are registered onto the left atrium 1’s endocardium mesh. Name left atrium 1’s endocardium the target mesh, other endocardium or epicardium mesh a moving mesh. First, rotate the moving mesh to roughly align with the target mesh. Next, apply a rigid ICP algorithm to optimize the alignment. Lastly, apply a non-rigid ICP algorithm to morph the moving mesh into the shape of the target mesh \cite{Amberg2007}. Fig. \ref{fig:fiber registration} shows an example of the mesh morphing. The blue mesh is the target mesh, the red mesh (can be endocardium or epicardium) is the moving mesh. (a) shows the original meshes. (b) shows the red mesh is morphed into the shape of the blue mesh. We can see that the morphed mesh matches the target well, on average, the distance between the nearest vertex pairs between the morphed mesh and the target mesh is: 1.4 +/- 0.5 mm. Then we need to morph the fiber data. (c) shows the process: it is a reference frame transformation. The final step is to register the morphed fiber to the target mesh, this can be done by copying the fibers that are nearest to the target mesh's triangles.

\section{Compare different fiber organizations}
\label{app:compare fiber}

\begin{figure}[!ht]
\centering
\includegraphics[width = 0.5\textwidth]{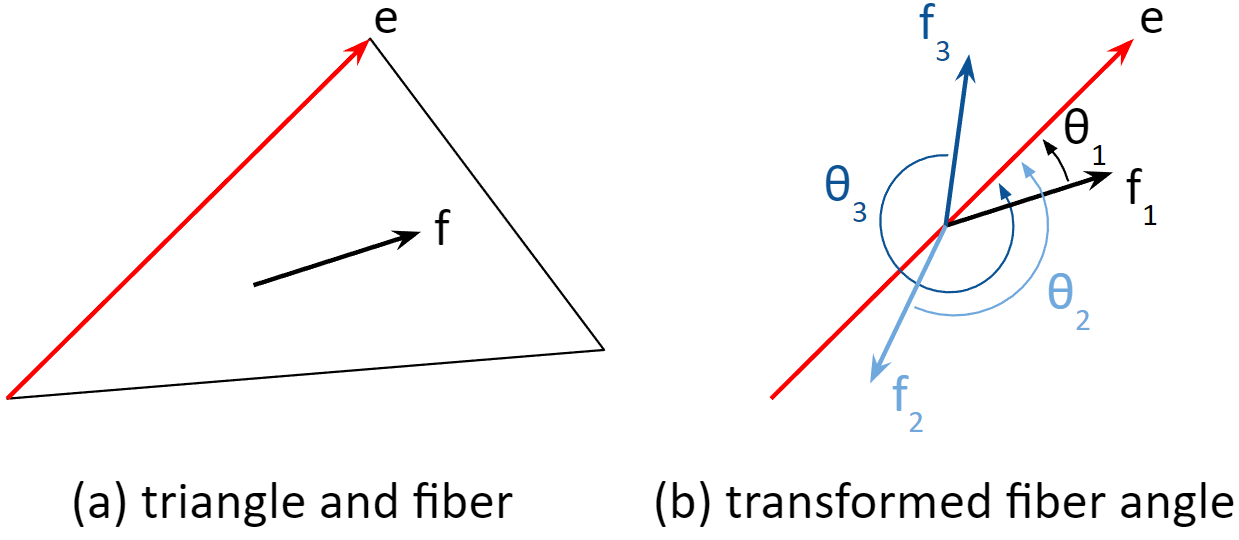}
\caption{Transform fiber data. (a) Fiber $f$ resides in a triangle of the mesh. The reference vector $e$ created from a fixed edge, all the seven different fibers will reference this vector $e$. (b) The 3D fiber vector is transformed into a 1D value. Examples of fibers ($f_1$, $f_2$, and $f_3$) and their transformed values ($\theta_1$, $\theta_2$, and $\theta_3$).}
\label{fig:compare fiber}
\end{figure}

There are seven different fiber organizations registered onto the left atrium 1’s endocardium mesh. Fibers are 3D vectors as shown in Fig. \ref{fig:compare fiber}(a), it is not convenient to directly compute correlations among 3D vectors. Therefore, we make a reference frame transformation so that the 3D fibers ($f_1$, $f_2$, and $f_3$) are represented as 1D values ($\theta_1$, $\theta_2$, and $\theta_3$) as shown in (b). The reference frame is a vector created by a fixed edge of each triangle (vector $e$ in (a)), and the transformation is to compute the angle between the fiber and that fixed edge ($\theta$ in (b)) follow the right hand rule of the triangle face normal vector. (Note that it is a property of a triangular mesh, that the sequence of the 3 vertices of a triangle are in such a way that we can utilize the right-hand-rule to find out the triangular face normal vector that points outwards of the mesh.) After this transformation, we can easily compute the correlations between different fiber organizations.

\section{Spatial and temporal resolutions}
\label{app:resolution}

\begin{figure}[!ht]
\centering
\includegraphics[width = 0.5\textwidth]{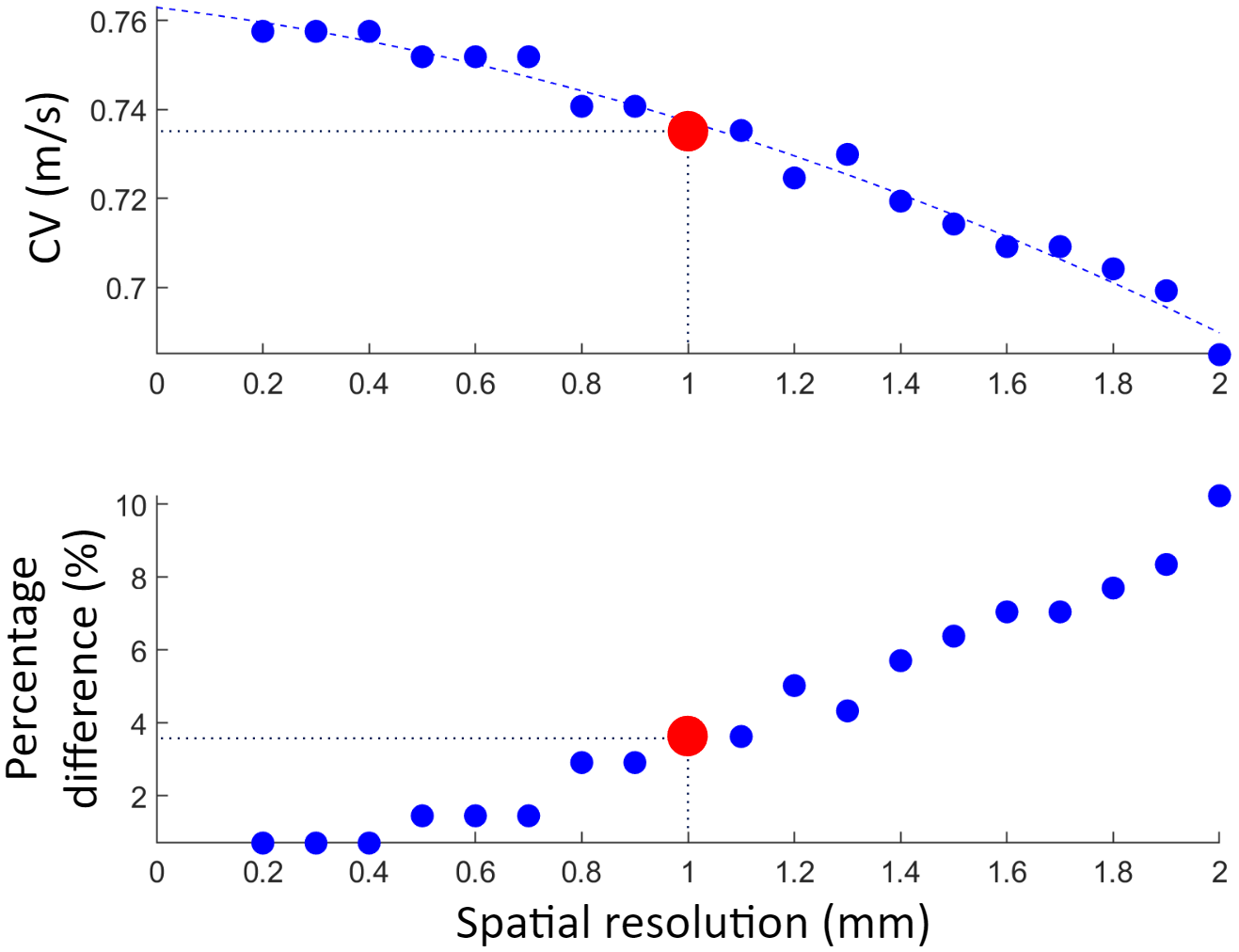}
\caption{Top: Conduction velocity (CV) values (the dots) of implementing different resolutions. The dashed blue line is a quadratic fit of the dots. Bottom: \ac{CV} percentage differences compare to the asymptotic value, which is the value of the dashed blue line at 0 mm spatial resolution. For spatial resolution of 1 mm and temporal resolution of 0.01 ms (the big red dot), the \ac{CV} percentage difference is 3.6\%, much smaller than the adequate 10\%.}
\label{fig:delta vs conduction velocity}
\end{figure}

We ran simulations on a slab of 100 mm $\times$ 4 mm $\times$ 4 mm. (The long length of the slab is to help increase \ac{CV} computational accuracy.) The heart model parameters are chosen such that the \ac{CV} values are close to the physical values \cite{Harrild2000}. Assume isotropic conduction, or no fiber organization. Spatial resolutions are set to 0.2, 0.3, ..., 2.0 mm. Temporal resolution $dt$ is set to 0.01 ms. Results are summarized in Fig. \ref{fig:delta vs conduction velocity}. In this paper, spatial resolution is 1 mm and temporal resolution is 0.01 ms. Using such resolutions, the accuracy of \ac{CV} is adequate with a deviation from the asymptotic value of 3.6\%, which is much smaller than the usual required 10\%.

\end{appendices}

\bibliographystyle{ieeetr}
\bibliography{reference}

\begin{thebibliography}{10}

\bibitem{o2011simulation}
T.~O'Hara {\em et~al.}, ``Simulation of the undiseased human cardiac
  ventricular action potential: model formulation and experimental
  validation,'' {\em PLoS computational biology}, vol.~7, no.~5, p.~e1002061,
  2011.

\bibitem{greene2022voltage}
D.~Greene {\em et~al.}, ``Voltage-mediated mechanism for calcium wave
  synchronization and arrhythmogenesis in atrial tissue,'' {\em Biophysical
  Journal}, vol.~121, no.~3, pp.~383--395, 2022.

\bibitem{fenton2008models}
F.~H. Fenton and E.~M. Cherry, ``Models of cardiac cell,'' {\em Scholarpedia},
  vol.~3, no.~8, p.~1868, 2008.

\bibitem{Ho1999}
S.~Ho {\em et~al.}, ``Anatomy of the left atrium: implications for
  radiofrequency ablation of atrial fibrillation,'' {\em Journal of
  cardiovascular electrophysiology}, vol.~10, p.~1525—1533, November 1999.

\bibitem{Ho2009}
Y.~Ho and D.~Sánchez-Quintana, ``The importance of atrial structure and
  fibers,'' {\em Clin Anat}, vol.~22, no.~1, pp.~52--63, 2009.

\bibitem{iaizzo2016visible}
P.~A. Iaizzo, ``The visible heart{\textregistered} project and free-access
  website ‘atlas of human cardiac anatomy’,'' {\em EP Europace}, vol.~18,
  no.~suppl\_4, pp.~iv163--iv172, 2016.

\bibitem{Ho2001}
S.~Ho {\em et~al.}, ``Architecture of the pulmonary veins: relevance to
  radiofrequency ablation,'' {\em Heart}, vol.~86, no.~3, pp.~265--270, 2001.

\bibitem{Fastl2018}
T.~E. Fastl {\em et~al.}, ``Personalized computational modeling of left atrial
  geometry and transmural myofiber architecture,'' {\em Medical Image
  Analysis}, vol.~47, pp.~180--190, 2018.

\bibitem{Lopez-Perez2015}
A.~Lopez-Perez {\em et~al.}, ``Three-dimensional cardiac computational
  modelling: methods, features and applications,'' {\em BioMedical Engineering
  OnLine}, vol.~14, p.~35, 2015.

\bibitem{berman2021interactive}
J.~P. Berman {\em et~al.}, ``Interactive 3d human heart simulations on
  segmented human mri hearts,'' in {\em 2021 Computing in Cardiology (CinC)},
  vol.~48, pp.~1--4, IEEE, 2021.

\bibitem{kaboudian2019real}
A.~Kaboudian {\em et~al.}, ``Real-time interactive simulations of large-scale
  systems on personal computers and cell phones: Toward patient-specific heart
  modeling and other applications,'' {\em Science advances}, vol.~5, no.~3,
  p.~eaav6019, 2019.

\bibitem{weiss2002electrical}
J.~N. Weiss {\em et~al.}, ``Electrical restitution and cardiac fibrillation,''
  {\em Journal of cardiovascular electrophysiology}, vol.~13, no.~3,
  pp.~292--295, 2002.

\bibitem{watanabe2001mechanisms}
M.~A. Watanabe {\em et~al.}, ``Mechanisms for discordant alternans,'' {\em
  Journal of cardiovascular electrophysiology}, vol.~12, no.~2, pp.~196--206,
  2001.

\bibitem{fenton1998fiber}
F.~Fenton and A.~Karma, ``Fiber-rotation-induced vortex turbulence in thick
  myocardium,'' {\em Physical review letters}, vol.~81, no.~2, p.~481, 1998.

\bibitem{groenendaal2014voltage}
W.~Groenendaal {\em et~al.}, ``Voltage and calcium dynamics both underlie
  cellular alternans in cardiac myocytes,'' {\em Biophysical journal},
  vol.~106, no.~10, pp.~2222--2232, 2014.

\bibitem{uzelac2017simultaneous}
I.~Uzelac {\em et~al.}, ``Simultaneous quantification of spatially discordant
  alternans in voltage and intracellular calcium in langendorff-perfused rabbit
  hearts and inconsistencies with models of cardiac action potentials and ca
  transients,'' {\em Frontiers in physiology}, vol.~8, p.~819, 2017.

\bibitem{arevalo2016arrhythmia}
H.~J. Arevalo {\em et~al.}, ``Arrhythmia risk stratification of patients after
  myocardial infarction using personalized heart models,'' {\em Nature
  communications}, vol.~7, no.~1, pp.~1--8, 2016.

\bibitem{kayvanpour2015towards}
E.~Kayvanpour {\em et~al.}, ``Towards personalized cardiology: multi-scale
  modeling of the failing heart,'' {\em PLoS One}, vol.~10, no.~7, p.~e0134869,
  2015.

\bibitem{niederer2020creation}
S.~Niederer {\em et~al.}, ``Creation and application of virtual patient cohorts
  of heart models,'' {\em Philosophical Transactions of the Royal Society A},
  vol.~378, no.~2173, p.~20190558, 2020.

\bibitem{Sanguinetti2003}
M.~C. Sanguinetti and P.~B. Bennett, ``Antiarrhythmic drug target choices and
  screening,'' {\em Circulation Research}, vol.~93, no.~6, pp.~491--499, 2003.

\bibitem{bai2020silico}
J.~Bai {\em et~al.}, ``In silico study of the effects of anti-arrhythmic drug
  treatment on sinoatrial node function for patients with atrial
  fibrillation,'' {\em Scientific reports}, vol.~10, no.~1, pp.~1--14, 2020.

\bibitem{uzelac2021quantifying}
I.~Uzelac {\em et~al.}, ``Quantifying arrhythmic long qt effects of
  hydroxychloroquine and azithromycin with whole-heart optical mapping and
  simulations,'' {\em Heart Rhythm O2}, vol.~2, no.~4, pp.~394--404, 2021.

\bibitem{mamoshina2021toward}
P.~Mamoshina {\em et~al.}, ``Toward a broader view of mechanisms of drug
  cardiotoxicity,'' {\em Cell Reports Medicine}, vol.~2, no.~3, p.~100216,
  2021.

\bibitem{Duncker2017}
D.~Duncker and C.~Veltmann, ``Optimizing antitachycardia pacing,'' {\em
  Circulation: Arrhythmia and Electrophysiology}, vol.~10, no.~9, p.~e005696,
  2017.

\bibitem{ji2017synchronization}
Y.~C. Ji {\em et~al.}, ``Synchronization as a mechanism for low-energy
  anti-fibrillation pacing,'' {\em Heart rhythm}, vol.~14, no.~8,
  pp.~1254--1262, 2017.

\bibitem{detal2022terminating}
N.~DeTal {\em et~al.}, ``Terminating spiral waves with a single designed
  stimulus: Teleportation as the mechanism for defibrillation,'' {\em
  Proceedings of the National Academy of Sciences}, vol.~119, no.~24,
  p.~e2117568119, 2022.

\bibitem{Lim2020}
B.~Lim {\em et~al.}, ``In situ procedure for high-efficiency computational
  modeling of atrial fibrillation reflecting personal anatomy, fiber
  orientation, fibrosis, and electrophysiology,'' {\em Scientific Reports},
  vol.~10, p.~2417, 2020.

\bibitem{prakosa2018personalized}
A.~Prakosa {\em et~al.}, ``Personalized virtual-heart technology for guiding
  the ablation of infarct-related ventricular tachycardia,'' {\em Nature
  biomedical engineering}, vol.~2, no.~10, pp.~732--740, 2018.

\bibitem{pathmanathan2019comprehensive}
P.~Pathmanathan {\em et~al.}, ``Comprehensive uncertainty quantification and
  sensitivity analysis for cardiac action potential models,'' {\em Frontiers in
  physiology}, vol.~10, p.~721, 2019.

\bibitem{pathmanathan2020data}
P.~Pathmanathan {\em et~al.}, ``Data-driven uncertainty quantification for
  cardiac electrophysiological models: Impact of physiological variability on
  action potential and spiral wave dynamics,'' {\em Frontiers in physiology},
  vol.~11, p.~585400, 2020.

\bibitem{doste2019rule}
R.~Doste {\em et~al.}, ``A rule-based method to model myocardial fiber
  orientation in cardiac biventricular geometries with outflow tracts,'' {\em
  International journal for numerical methods in biomedical engineering},
  vol.~35, no.~4, p.~e3185, 2019.

\bibitem{bayer2012novel}
J.~D. Bayer {\em et~al.}, ``A novel rule-based algorithm for assigning
  myocardial fiber orientation to computational heart models,'' {\em Annals of
  biomedical engineering}, vol.~40, no.~10, pp.~2243--2254, 2012.

\bibitem{Valderrabano2007}
M.~Valderrábano, ``Influence of anisotropic conduction properties in the
  propagation of the cardiac action potential,'' {\em Progress in Biophysics
  and Molecular Biology}, vol.~94, no.~1, pp.~144--168, 2007.
\newblock Gap junction channels: from protein genes to diseases.

\bibitem{franzone1993spread}
P.~C. Franzone {\em et~al.}, ``Spread of excitation in a myocardial volume:
  Simulation studies in a model of anisotropic ventricular muscle activated by
  point stimulation,'' {\em Journal of cardiovascular electrophysiology},
  vol.~4, no.~2, pp.~144--160, 1993.

\bibitem{fenton1998vortex}
F.~Fenton and A.~Karma, ``Vortex dynamics in three-dimensional continuous
  myocardium with fiber rotation: Filament instability and fibrillation,'' {\em
  Chaos: An Interdisciplinary Journal of Nonlinear Science}, vol.~8, no.~1,
  pp.~20--47, 1998.

\bibitem{Bhakta2008}
D.~Bhakta and J.~Miller, ``Principles of electroanatomic mapping,'' {\em Indian
  Pacing Electrophysiol J.}, vol.~1, no.~8, pp.~32--50, 2008.

\bibitem{Zhao2012}
J.~Zhao {\em et~al.}, ``An image-based model of atrial muscular architecture,''
  {\em Circulation: Arrhythmia and Electrophysiology}, vol.~5, no.~2,
  pp.~361--370, 2012.

\bibitem{Pashakhanloo2016}
F.~Pashakhanloo {\em et~al.}, ``Myofiber architecture of the human atria as
  revealed by submillimeter diffusion tensor imaging,'' {\em Circulation:
  Arrhythmia and Electrophysiology}, vol.~9, no.~4, p.~e004133, 2016.

\bibitem{Roney2021}
C.~H. Roney {\em et~al.}, ``Constructing a human atrial fibre atlas,'' {\em
  Annals of Biomedical Engineering}, vol.~49, p.~233–250, 2021.

\bibitem{Whitaker2016}
J.~Whitaker {\em et~al.}, ``{The role of myocardial wall thickness in atrial
  arrhythmogenesis},'' {\em EP Europace}, vol.~18, pp.~1758--1772, 05 2016.

\bibitem{Sun2018}
J.~Y. Sun {\em et~al.}, ``Left atrium wall-mapping application for wall
  thickness visualisation,'' {\em Sci Rep}, vol.~8, p.~4169, 2018.

\bibitem{Mitchell2003}
C.~C. Mitchell and D.~G. Schaeffer, ``A two-current model for the dynamics of
  cardiac membrane,'' {\em Bulletin of Mathematical Biology}, vol.~65,
  p.~767–793, 2003.

\bibitem{Elaff2018}
I.~Elaff, ``Modeling of realistic heart electrical excitation based on dti
  scans and modified reaction diffusion equation,'' {\em Turkish Journal of
  Electrical Engineering \& Computer Sciences}, vol.~26, no.~3, pp.~1153--1163,
  2018.

\bibitem{McFarlane2010}
R.~McFarlane, ``High-performance computing for computational biology of the
  heart,'' {\em Doctor in Philosophy thesis of the University of Liverpool,
  School of Electrical Engineering, Electronics and Computer Science}, p.~132,
  2010.

\bibitem{Cabrera2017}
R.~Cabrera-Lozoya {\em et~al.}, ``Image-based biophysical simulation of
  intracardiac abnormal ventricular electrograms,'' {\em IEEE Transactions on
  Biomedical Engineering}, vol.~64, no.~7, pp.~1446--1454, 2017.

\bibitem{Roney2019}
C.~H. Roney {\em et~al.}, ``A technique for measuring anisotropy in atrial
  conduction to estimate conduction velocity and atrial fibre direction,'' {\em
  Computers in Biology and Medicine}, vol.~104, pp.~278--290, 2019.

\bibitem{Coster2018}
T.~D. Coster {\em et~al.}, ``Myocyte remodeling due to fibro-fatty
  infiltrations influences arrhythmogenicity,'' {\em Front. Physiol.}, vol.~9,
  p.~1381, 2018.

\bibitem{Aslanidi2011}
O.~V. Aslanidi {\em et~al.}, ``3d virtual human atria: A computational platform
  for studying clinical atrial fibrillation,'' {\em Prog Biophys Mol Biol},
  vol.~107(1), pp.~156--68, 2011.

\bibitem{Boyle2019}
P.~M. Boyle {\em et~al.}, ``Computationally guided personalized targeted
  ablation of persistent atrial fibrillation,'' {\em Nature Biomedical
  Engineering}, vol.~3, p.~870–879, 2019.

\bibitem{Ho2012}
Y.~Ho {\em et~al.}, ``Left atrial anatomy revisited,'' {\em Circulation:
  Arrhythmia and Electrophysiology}, vol.~5, no.~1, pp.~220--228, 2012.

\bibitem{Krueger2011}
M.~W. Krueger {\em et~al.}, ``Modeling atrial fiber orientation in
  patient-specific geometries: A semi-automatic rule-based approach,'' in {\em
  Springer Berlin Heidelberg}, pp.~223--232, 2011.

\bibitem{Labarthe2021}
S.~Labarthe {\em et~al.}, ``A semi-automatic method to construct atrial fibre
  structures: A tool for atrial simulations,'' in {\em 2012 Computing in
  Cardiology}, pp.~881--884, 2012.

\bibitem{Wachter2015}
A.~Wachter {\em et~al.}, ``Mesh structure-independent modeling of
  patient-specific atrial fiber orientation,'' {\em Current Directions in
  Biomedical Engineering}, vol.~1, no.~1, pp.~409--412, 2015.

\bibitem{Saliani2021}
A.~Saliani {\em et~al.}, ``Simulation of diffuse and stringy fibrosis in a
  bilayer interconnected cable model of the left atrium,'' {\em EP Europace},
  vol.~23, no.~Supplement-1, pp.~i169--i177, 2021.

\bibitem{Cantwell2015TechniquesFA}
C.~D. Cantwell {\em et~al.}, ``Techniques for automated local activation time
  annotation and conduction velocity estimation in cardiac mapping,'' {\em
  Computers in Biology and Medicine}, vol.~65, pp.~229 -- 242, 2015.

\bibitem{He2021}
J.~He {\em et~al.}, ``Patient-specific heart model towards atrial
  fibrillation,'' in {\em Proceedings of the ACM/IEEE 12th International
  Conference on Cyber-Physical Systems}, ICCPS '21, (New York, NY, USA),
  p.~33–43, Association for Computing Machinery, 2021.

\bibitem{He2019}
J.~He {\em et~al.}, ``Electroanatomic mapping to determine scar regions in
  patients with atrial fibrillation,'' in {\em 2019 41st Annual International
  Conference of the IEEE Engineering in Medicine and Biology Society (EMBC)},
  pp.~5941--5944, 2019.

\bibitem{Virag2002}
N.~Virag {\em et~al.}, ``Study of atrial arrhythmias in a computer model based
  on magnetic resonance images of human atria,'' {\em Chaos}, vol.~12, no.~3,
  pp.~754--763, 2002.

\bibitem{Ruchat2007}
R.~P {\em et~al.}, ``A biophysical model of atrial fibrillation ablation: what
  can a surgeon learn from a computer model?,'' {\em Europace}, vol.~9, no.~6,
  pp.~vi71--6, 2007.

\bibitem{Gray1996}
G.~RA {\em et~al.}, ``Incomplete reentry and epicardial breakthrough patterns
  during atrial fibrillation in the sheep heart,'' {\em Circulation}, vol.~94,
  no.~10, pp.~2649--61, 1996.

\bibitem{Gonzales2014}
M.~Gonzales {\em et~al.}, ``Structural contributions to fibrillatory rotors in
  a patient-derived computational model of the atria,'' {\em Europace},
  pp.~iv3--iv10, 2014.

\bibitem{Roth2021}
B.~J. Roth, ``Bidomain modeling of electrical and mechanical properties of
  cardiac tissue,'' {\em Biophysics Reviews}, vol.~2, no.~4, p.~041301, 2021.

\bibitem{Potse2006}
M.~Potse {\em et~al.}, ``A comparison of monodomain and bidomain
  reaction-diffusion models for action potential propagation in the human
  heart,'' {\em IEEE Transactions on Biomedical Engineering}, vol.~53, no.~12,
  pp.~2425--2435, 2006.

\bibitem{de2016direct}
N.~de~Groot {\em et~al.}, ``Direct proof of endo-epicardial asynchrony of the
  atrial wall during atrial fibrillation in humans,'' {\em Circulation:
  Arrhythmia and Electrophysiology}, vol.~9, no.~5, p.~e003648, 2016.

\bibitem{eckstein2013transmural}
J.~Eckstein {\em et~al.}, ``Transmural conduction is the predominant mechanism
  of breakthrough during atrial fibrillation: evidence from simultaneous
  endo-epicardial high-density activation mapping,'' {\em Circulation:
  Arrhythmia and Electrophysiology}, vol.~6, no.~2, pp.~334--341, 2013.

\bibitem{verheule2014role}
S.~Verheule {\em et~al.}, ``Role of endo-epicardial dissociation of electrical
  activity and transmural conduction in the development of persistent atrial
  fibrillation,'' {\em Progress in biophysics and molecular biology}, vol.~115,
  no.~2-3, pp.~173--185, 2014.

\bibitem{wu1998role}
T.-J. Wu {\em et~al.}, ``Role of pectinate muscle bundles in the generation and
  maintenance of intra-atrial reentry: potential implications for the mechanism
  of conversion between atrial fibrillation and atrial flutter,'' {\em
  Circulation research}, vol.~83, no.~4, pp.~448--462, 1998.

\bibitem{Amberg2007}
B.~Amberg {\em et~al.}, ``Optimal step nonrigid icp algorithms for surface
  registration,'' in {\em 2007 IEEE Conference on Computer Vision and Pattern
  Recognition}, pp.~1--8, 2007.

\bibitem{Harrild2000}
D.~M. Harrild and C.~S. Henriquez, ``A computer model of normal conduction in
  the human atria,'' {\em Circulation Research}, vol.~87, no.~7, pp.~e25--e36,
  2000.

\end{thebibliography}

\end{document}